\pdfoutput=1

\documentclass[
twocolumn,times,tighten,twocolappendix, 
]{aastex631}
\usepackage{graphicx,graphics,amsmath}
\usepackage{natbib}
\usepackage{bm,url}
\usepackage{color}
\usepackage{array} % need to have this just before `arydshln` package, to remove `hrule` errors
\usepackage{multirow}
\usepackage{hyperref}
\usepackage{float}
\usepackage{csvsimple}
\usepackage{booktabs} % for `hline`

\graphicspath{{./figs/}{./png/}}

\def\bl{Babcock--Leighton}
\newcommand{\Fig}[1]{Figure~\ref{#1}}

\newcommand{\Sec}[1]{Section~\ref{#1}}
\newcommand{\Tab}[1]{Table~\ref{#1}}

\emergencystretch=\maxdimen
\shorttitle{Role of sunspot latitude versus tilt in determining the cycle amplitude}
\shortauthors{Dey et al.}

\begin{document}

\title{Role of sunspot latitude versus tilt in determining the polar field and amplitude of the next cycle: Cause of the weak Solar Cycle 20}

\author{Bidisha Dey}
\email{bidisha\_20221030@students.iisertirupati.ac.in}
\affil{Indian Institute of Science Education and Research, Tirupati 517619, India}

\author[0000-0001-7036-2902]{Anu Sreedevi}
\email{anubsreedevi.rs.phy20@itbhu.ac.in}
\affil{Department of Physics, Indian Institute of Technology (Banaras Hindu University), Varanasi 221005, India}

\author[0000-0002-8883-3562]{Bidya Binay Karak} 
\email{karak.app@iitbhu.ac.in}
\affil{Department of Physics, Indian Institute of Technology (Banaras Hindu University), Varanasi 221005, India}
\correspondingauthor{Bidya Binay Karak}
\email{karak.app@iitbhu.ac.in}

% \email{karak.app@iitbhu.ac.in}

\begin{abstract}
One prominent feature of solar cycle is its irregular variation in its cycle strength, making it challenging to predict the amplitude of the next cycle. Studies show that fluctuations and nonlinearity in the process of generating poloidal field through the decay and dispersal of tilted sunspots produce variation in the solar cycle. The flux, latitudinal position, and tilt angle of sunspots are the primary parameters that determine the polar field and, thus, the next solar cycle strength. 
By analyzing the observed sunspots and polar field proxy, we show that the nonlinearity in the poloidal field generation becomes important for strong cycles. Except for strong cycles, we can reasonably predict the polar field at the end of the cycle (and thus the next cycle strength) using the total sunspot area alone. Combining the mean tilt angle and latitude positions with the sunspot area, we can predict the polar field of Cycles 15---24 (or the amplitude of sunspot Cycles 16--25) with reasonable accuracy, except for Cycle 23 for which the average tilt angle cannot predict the polar field. For Cycles 15--22, we show that the average tilt angle variation dominates over the latitude variation in determining the polar field of a cycle. In particular, the reduction of tilt in Cycle 19 was the primary cause of the following weak cycle (Cycle 20). Thus, we conclude that tilt quenching is essential in regulating the solar cycle strength in the solar dynamo.  
\end{abstract}
%\keywords{Solar dynamo, Sunspots, Magnetic fields, Solar Cycle, Polar field, solar cycle prediction}

%\end{frontmatter}
%-------------------------------------------------

\section{Introduction}
     \label{S-Introduction} 
The number of sunspots or the strength of the magnetic field observed on the solar surface varies cyclically with an average period of 11 years, forming a popular cycle of nature, known as the sunspot or solar cycle.  A notable feature of the solar cycle is that not all cycles are identical. In particular, the cycle amplitude exhibits considerable irregularity from one cycle to another, making it challenging to predict the next cycle \citep{Petrovay20}. Beyond the 11-year periodicity, the solar cycle also displays other variations, ranging from short-term fluctuations on the scale of days (such as the Rieger periodicity), to long-term modulations like the ~90-year Gleissberg cycle, and even extended periods of exceptionally low or high activity, such as the Maunder Minimum or Grand Maxima \citep{glessberg, Biswas23}. These variations have significant impacts on Earth’s climate and space climate, and the 11-year cycle plays a particularly central role. Therefore, understanding the full spectrum of solar variability and improving our ability to predict it remain a key scientific goal. 

The cyclic appearance of sunspots is driven by a dynamo  operating in the Sun's convection zone \citep{Pa55}. Dynamo is a process in which Sun's magnetic field configuration loops between poloidal and toroidal. Here, the toroidal field is generated by the shearing of the poloidal field by differential rotation in the convection zone (the so-called $\Omega$ effect). Due to magnetic buoyancy, the toroidal flux (tubes) rise to form tilted bipolar magnetic regions (BMRs or sunspots as observed in white light) \citep{Pa55b}. During the rise of the toroidal flux inside the convection zone, the tilt is presumably produced by the torque induced on the diverging flows arising from the apex of the magnetic flux tube \citep{DC93}. These tilted BMRs decay and produce a poloidal field, popularly known as the \bl\ process \citep{Ba61, Le64}, which acts as the seed for the next cycle and completes the dynamo loop. Observations, complemented by theoretical investigations, suggest that this \bl\ process is a dominating mechanism for generating the poloidal field in the Sun \citep{CS15}. As the $\Omega$ effect is (largely) linear, we observe a linear correlation between the poloidal field at solar minimum and the toroidal field of the next cycle. In observations, we also observe a strong correlation between the peaks (or the growth rate) of the polar field and the amplitude of the next sunspot cycle \citep{WS09, MMS89, Muno13, kumar21, kumar22}. 
This provides a physical basis for the solar cycle prediction \citep{Sch78, CCJ07, Petrovay20}. Interestingly, this correlation is not limited to the \bl\ type dynamo models, but rather robustly holds in any dynamo model as long as the poloidal field is fed to the shear layer \citep{CB11, Kumar21b}.

\begin{figure*}    %%%%%%%%%%%%%%%%%% FIGURE 1 
\centerline{\includegraphics[width=\textwidth]{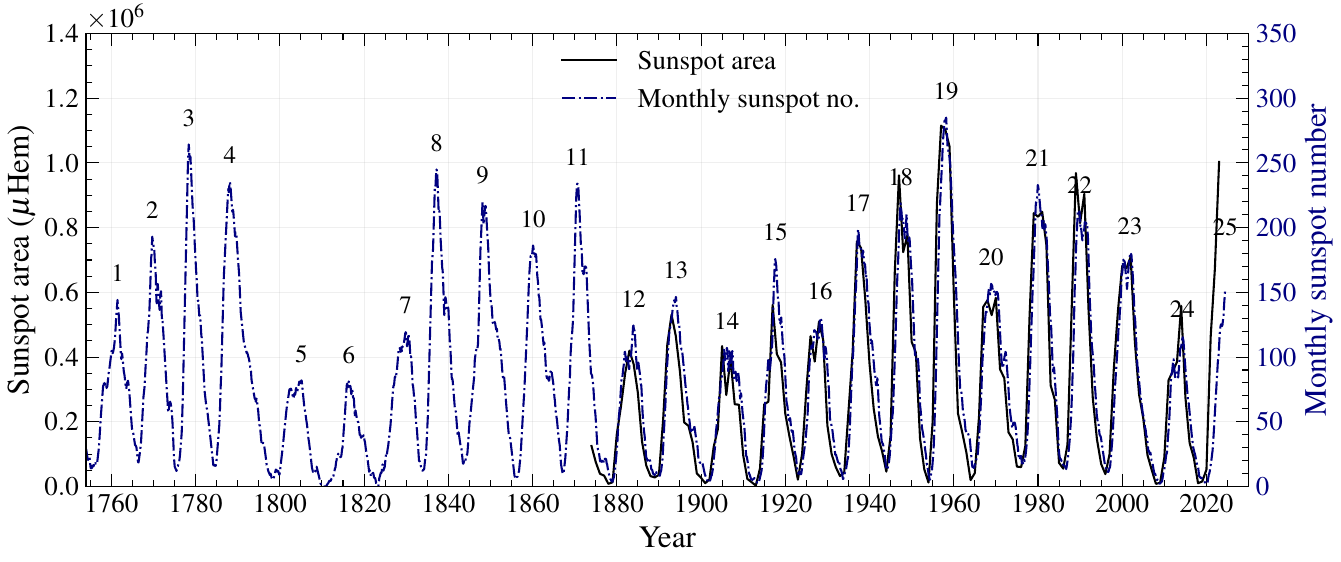}}
\small
        \caption{Solid black line represents the daily calibrated group 
        area as a function of time from \citet{Mandal2020}. 
        Dashed red line represents the monthly total sunspot number (International Sunspot Number V2) from WDC-SILSO, Royal Observatory of Belgium, Brussels. The cycle numbers are marked on the plot. 
        }
        
\label{fig:ts}
\end{figure*}

Now, if the dynamo is just a simple loop of toroidal and poloidal field conversions, then why do we observe such a significant variation in the cycle amplitude? The answer for this question lies in the stochastic fluctuations in the sunspot emergence caused by convective turbulence and nonlinearities  \citep{Karak23}. There is ample evidence that fluctuations in the BMR properties, namely, the tilt, flux, time delay of emergence and latitude, as observed by their distributions, play major roles in disturbing the generation of the poloidal field \citep{Kumar24}. On top of these, the nonlinearity adds additional complexity to making the polar field unequal \citep{Talafha22}. Two potential nonlinearities have been identified in the \bl\ framework: tilt quenching \citep{LC17, KM17} and latitude quenching \citep{Petrovay20, J20, Kar20}.  
Based on the thin flux tube model, which explains the formation of tilt as given by Joy's law, it is predicted that a strong magnetic flux tube rises rapidly, giving less time to the Coriolis force to produce tilt. Therefore, strong BMRs should have less tilt \citep{FFM94}. There was some evidence for it in terms of the individual BMR tilt with magnetic field strength measured from magnetograms of the last two solar cycles \citep{jha20, Sreedevi24} and in the cycle average mean tilt vs the cycle strength in the white light data \citep{Das10, Das13, Jiao21}. Thus, based on this tilt quenching, we expect a strong cycle to produce a weak polar field, and thus the weak next cycle.   Additionally, it was found in observations that a strong cycle produces more BMRs at high latitude \citep{W55, SWS08,  Mandal17, BKC22}, which are less efficient in generating a polar field due to inefficient cross-equatorial cancellation of the leading polarity flux at the equator, as demonstrated in models \citep{KM18, J20}. Thus, similar to tilt quenching, we expect that a strong cycle should produce a weak polar field. This is the essence of latitude quenching. In the present article, we shall check to what extent these two quenchings hold and their roles in determining the polar field (or next sunspot cycle strength) in the data of past cycles. We shall also explore the role of stochastic fluctuations in determining the polar field strength. Finally, we shall explore the role of nonlinearity in determining the amplitude of the Cycle 19--20 pair, where the amplitude abruptly drops by a considerable amount after a strong cycle; \Fig{fig:ts}.

\section{Results}
\label{sec:res}

\subsection{Signature of nonlinearity and stochasticity in the solar cycle}

\begin{figure}%[!tbp]
  %\centering
  \vspace{1.06em}
    \includegraphics[width=0.48\textwidth]{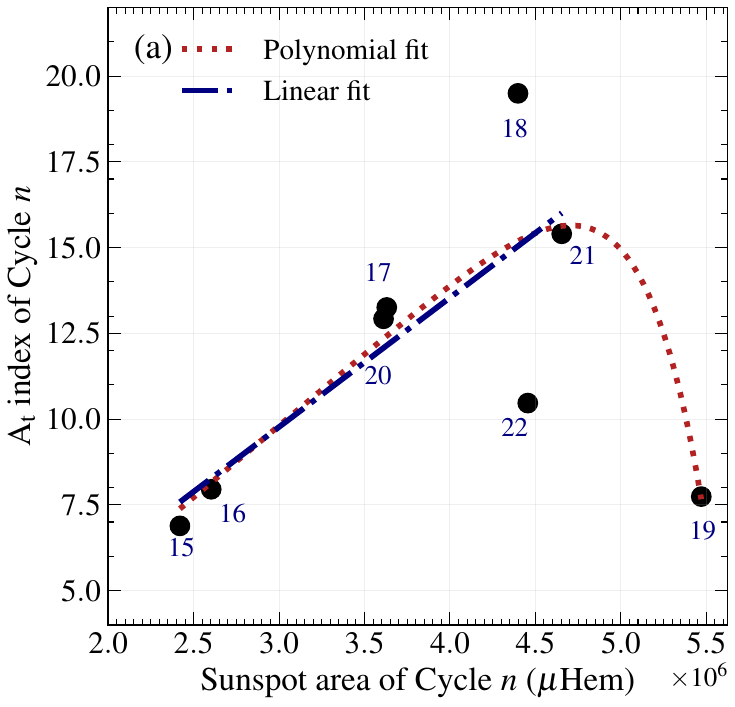}\label{fig:f1}
    
   % \vspace{5.45em}
    \includegraphics[width =0.48\textwidth,height = 0.48\textwidth]{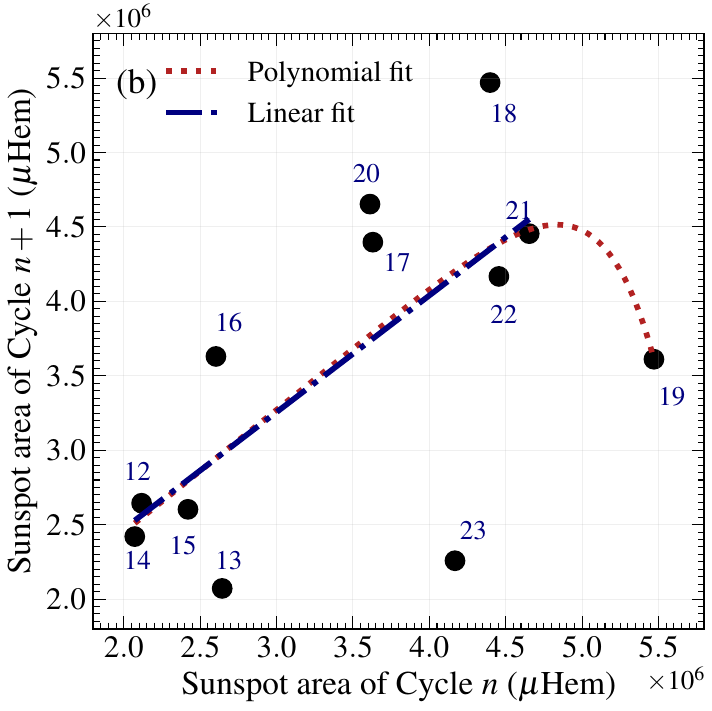}\label{fig:f2}
    \caption{(a) Polar field strength ($A(t)$ index) at the solar minimum plotted against the total sunspot area (strength) of the same solar cycle for Cycles 15--22.
    (b) Same as (a), but the amplitude of the next solar cycle as measured by the total sunspot area for Cycles 12--23. The cycle numbers are labeled to correspond to the value of $n$ (horizontal axis). The red dashed line shows a polynomial fit, while the blue solid line represents a linear fit excluding the data point for Cycle 19.
}
  \label{fig:quench}
\end{figure}

In the \bl\ process, the decay and dispersal of  BMRs produce the polar field, which determines the amplitude of the next sunspot cycle. So at a fundamental level, 
we can expect that if a cycle produces more BMRs, i.e., if a cycle is strong, then the generated polar field at the end of the cycle will be high. Indeed, in \Fig{fig:quench}(a), we observe a fairly reasonable linear trend between the total area (a measure of the flux) of the sunspots in a cycle and the polar field at the end of the same cycle. We note that the sunspot area is from the white-light observations from  \citet{Mandal2020} and the $A(t)$ index of \citet{Makarov01} is taken as a measure of the polar field strength. The $A(t)$ index is indeed a good representation of the polar field as found in several previous studies \citep{kumar21}, although it was obtained by taking only the dipole and octupole components of the magnetic field, reconstructed using the H$\alpha$ synoptic maps from Pulkovo Observatory. Here we have  used only Cycles 15--22 due to the availability of $A(t)$ index data.  In the next section, we shall use the WSO measurement of the polar field to extend this analysis for two more cycles. 
However, as the polar field strongly correlates with the amplitude of the next sunspot cycle, which is our target, we can make this correlation with the amplitude of the next sunspot cycle instead of the polar field. This, for cycles 12--24, is shown in \Fig{fig:quench}(b). Again, we observe a similar trend, as expected. 

Now we explain why the data points deviate from the linear trend in \Fig{fig:quench}(a) or (b). The first data point that draws our attention is for Cycle 19--20 pair, for which we observe a much weaker polar field than expected from the linear relation of the polar field with total area/flux of the sunspots. This is clearly a signature of the nonlinear quenching in the generation of the polar field. As discussed in the Introduction and detailed in \citet{J20} and \citet{Kar20}, both latitude and tilt quenchings can reduce the polar field. A strong cycle produces BMRs at high latitudes that are less efficient in generating the polar field, and the mean tilt of the BMR is also low in a strong cycle; both of these make the polar field or the amplitude of the next cycle weak. To identify which of these, the tilt or latitude, plays a dominant role for the Cycle 19--20 pair requires a detailed investigation of the data, which we do in \Sec{sec:subsec2}. 

\begin{figure}%[!tbp]
  \centering%\hfill
  \includegraphics[width=0.468\textwidth, height = 0.48\textwidth]{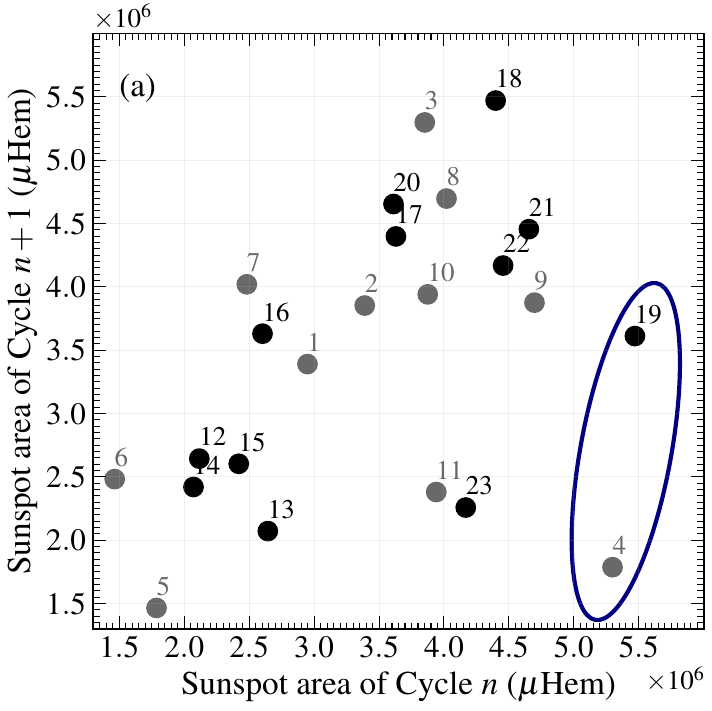}
  
\vspace{0.3cm}
  \includegraphics[width=0.48\textwidth, height = 0.463\textwidth]{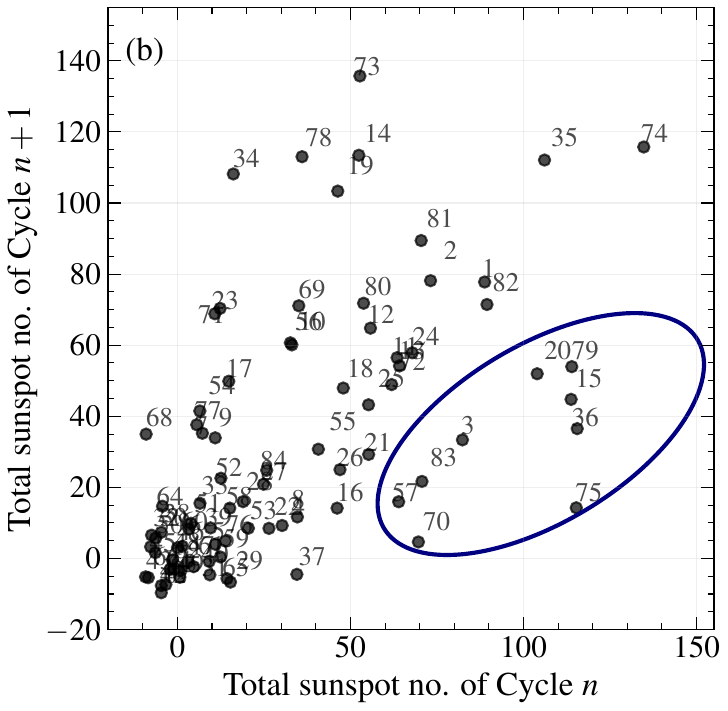}
  \caption{(a) Strength of cycle $n+1$ versus the strength of cycle $n$ as obtained from the sunspot area data. Note, the sunspot area data is available only from Cycle 12 onward. For earlier times, sunspot number data (V2) have been rescaled to match the sunspot area scale, and these points are shown in gray.
  (b) The same as (a) but for the reconstructed sunspot number from C$^{14}$ data of \citet{Usoskin21} and it is for the cycle average number and not the total number. Here we have 85 cycles and we follow the same convention for numbering the cycles as given in \citet{Usoskin21}. The oval highlights the data points for which the cycle strength of $n+1$ is significantly lower than that of cycle $n$. 
  }
 \label{fig:quench2}
\end{figure}

It may seem illogical that we are establishing our idea of observed nonlinear quenching based on one data point (Cycle 19--20 pair) in \Fig{fig:quench}. We do not have the polar field information earlier than Cycle 15 (although we have the polar proxy from polar active networks; \citet{mishra25} and faculae count from Cycle 14 to Cycle 23, the quality is poor\footnote{In fact, the physically expected correlation between the peak polar proxy around the solar minima and the strength (total area) of the next sunspot cycle is only 0.48 and 0.63 for the north and south polar faculae data respectively, while it is 0.92 for the $A(t)$  index.}) 
and sunspot area (a measure of BMR flux) beyond cycle 12. However, assuming the total sunspot numbers as a measure of total flux of BMRs in a cycle, we can extend the analysis for the last 24 cycles for which we have a fairly good record of sunspot numbers. \Fig{fig:quench2}(a) presents the strength of cycle $n+1$ (which is a measure of the polar field of cycle $n$) versus the strength of cycle $n$, as measured by the total number of sunspots in cycle $n$.  
%The plot appears messy, 
In this plot, the data points are more scattered, 
and the Cycle 19--20 data point has moved slightly towards the linear trend. However, we have an additional data point, which strengthens the quenching: the Cycle 4--5 pair. As seen in \Fig{fig:ts}, cycle~4 was another strong cycle, which caused the next cycle to be very weak. So, with limited data and reasonable assumptions, we can realize the existence of nonlinear quenching in generating the polar field and thus in the solar dynamo. With this reasoning, we can even more ambitiously look for the nonlinear trend in reconstructed sunspot number data (cycle average), which is available for the last 85 cycles \citep{Usoskin21}. In \Fig{fig:quench2}(b) we observe that there are points for which the amplitude of the next cycle is much reduced; see the data points inside the oval. We note that the solar cycle numbers that we know of does not correspond to the same here. Keeping in mind that the reconstructed sunspot data contain a huge amount of uncertainty \citep{Usoskin22}, we can identify a signature of nonlinear quenching in the parameter regime of cycle strength $n$ versus the cycle strength $n+1$ (a measure of the polar field strength of $n+1$). 

Next, we focus on the linear part of \Fig{fig:quench}. In the \bl\ process at first order, we expect that the polar field strength generated in a cycle should increase linearly with the increase of the total flux of the BMRs as long as the cycle is not too strong to reach the nonlinear quenching regime; also see Fig.~6 of \citet{J20}. However, in \Fig{fig:quench}, we observe a considerable scatter around the linear trend.  This is due to the random variations in the generation of the poloidal field through the \bl\ process.  In observations, it is seen that the BMR properties, namely the tilt angle (including anti-Hale and anti-Joy orientations), time delay of emergence, flux, and latitude, are not fixed but follow distributions which also vary cycle to cycle \citep{sk12, KM17, Sreedevi23, Kumar24}. Due to these varying properties of BMRs, the polar field generated at the end of a cycle differs from the expected value based on the total flux of BMRs \citep[e.g.,][]{Ca13, JCS14, Nagy17, KM17, pal23, BKK23, Kumar24}.

%===============================================================  
\subsection{Role of tilt quenching versus the latitude quenching}
\label{sec:subsec2}
Now we try to identify the parameters that determine the polar field strength (or the strength of the next cycle) for the past cycles. For this exploration, we shall first consider only cycles 15--22 because for these cycles, we have the data for the tilt angles of sunspots and uniform polar field proxy.

\begin{figure}%[!tbp]
  \centering
  \hfill
  \includegraphics[width=0.432\textwidth]{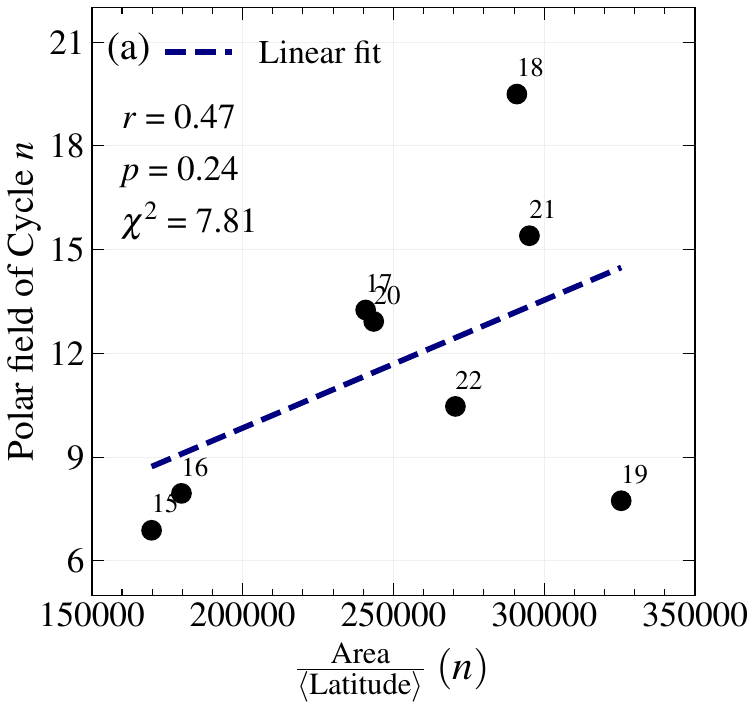}
 \hfill
  \includegraphics[width =0.42\textwidth]{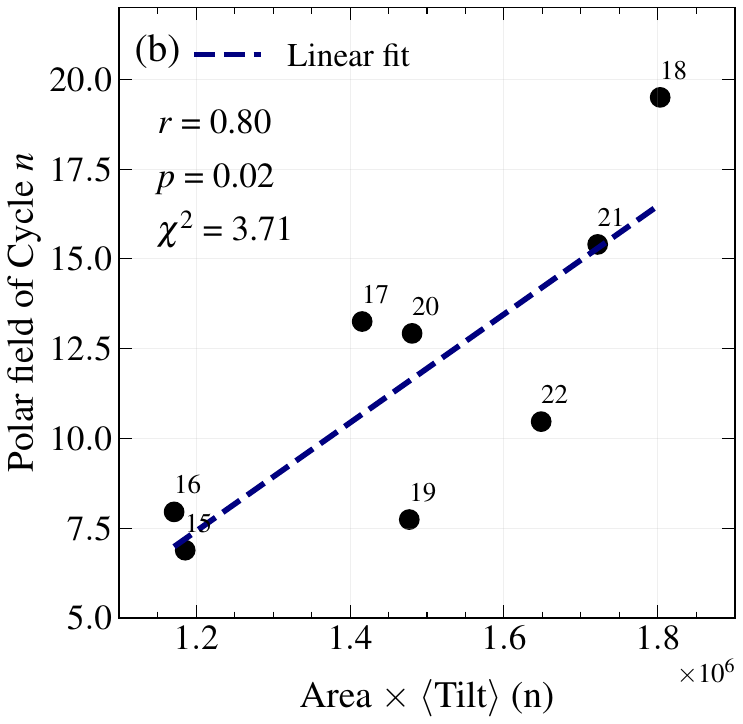}
  \hfill
  \includegraphics[width =0.42\textwidth]{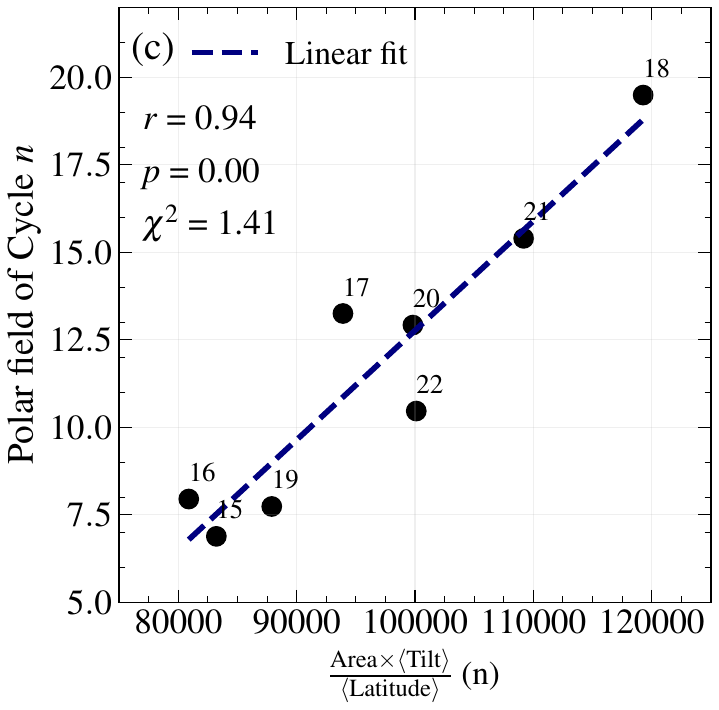}
  \caption{
  Polar field strength ($A(t)$ index) at the solar minimum vs the (a) total area divided by the mean latitude of sunspots, (b) total area times the mean tilt angle of sunspots, (c) total area times the mean tilt angle of sunspots divided the mean latitude of sunspots of a cycle. 
  %The corresponding linear (Pearson) correlation ($r$), $p$ and $\chi^2$ values are printed on each subplot. 
  }
  \label{fig:latvstilt}
\end{figure}

By now, it is evident that the polar field strength at the end of a cycle (or the amplitude of the next sunspot cycle) is primarily determined by the (i) total flux (sunspot area in the present case), (ii) average latitudinal positions, and  (iii)  average tilt angle of sunspots of the cycle. 
Referring to \Tab{table1} and \Fig{fig:quench}(a), we find that there is a weak linear correlation between the polar field at the minimum of the cycle $n$ or the strength of the cycle $n+1$ versus the total area of cycle $n$. There, the Cycle 19--20 pair was a prominent outlier, which in the previous section we identify as an evidence of nonlinear quenching in the polar field generation. 
Interestingly, if we exclude the data for this pair, the correlation significantly improves; see the $r^{-}_{19/20}$ value in \Tab{table1}. This is true for both the polar field and the strength of the next cycle. This implies that, except for the cycle 19--20 pair, the total sunspot area (a measure of the BMR flux) alone is a reasonable predictor of the polar field or the next cycle.  For the 19--20 pair, where a weak cycle was followed by a strong Cycle 19, additional physics is required. 
To identify this, we first capture the contribution of the sunspot latitudinal position. We know that a strong cycle produces sunspots at high latitudes, which tend to reduce the polar field. Hence, the generated polar field is expected to be correlated inversely with the mean sunspot latitude. So we divide the total area by the mean latitude of the sunspot over a cycle, and find its correlation with the polar field or the next sunspot cycle strength. However, now we observe only a slight improvement in the linear correlations (from 0.35 to 0.47 for polar field and 0.45 to 0.56 in sunspot area; see \Tab{table1} and \Fig{fig:latvstilt}(a)).  And again, the cycle 19--20 pair remains an outlier; excluding this point considerably increases the correlation. This suggests that while the mean latitude plays a role in determining the polar field at the end of the cycle, it does not explain why solar cycle 20 is so weak. This also raises doubts about the importance of latitude quenching in providing nonlinearity to stabilise the solar dynamo. It was thought that a strong cycle, which produces high-latitude spots, reduces the polar field, thus producing a weak following cycle \citep{J20}. However, this theoretical concept does not fully explain the sudden weak Cycle 20 after a very strong Cycle 19. 

\begin{table}
\begin{center}
\caption{
Linear (Pearson) correlation coefficients between the polar field strength and the total area (amplitude) of the next cycle with the total sunspot area (strength) of the present cycle. 
}
\begin{tabular}{cllcl}     % define the column alignment
                           % l: left, c: center, r: right
\hline                     % horizontal line
Cycles & \multicolumn{2}{c}{Quantities} & $r_{\rm all}$ & $r^{-}_{19/20}$\\
\hline                     % horizontal line
15--22 & Area ($n$) & Polar field ($n$) & 0.35$^{~}$& 0.78$^\ast$ \\
15--22 & ~~~ ---         & Area ($n+1$)      & 0.45$^{~}$& 0.77$^\ast$\\
\hline
15--22 & $\frac{ {\rm Area} }{ \langle {\rm Latitude} \rangle}$ ($n$)& Polar field ($n$)& 0.47$^{~}$&0.86$^\ast$\\
15--22 & ~~~ ---        & Area ($n$+1)& 0.56$^{~}$&0.84$^{*}$\\
 \hline
15--22 & ${\rm Area} \times \langle {\rm Tilt}\rangle$ ($n$)& Polar field ($n$)& 0.80$^\ast$&0.86$^\ast$\\
15--22  & ~~~ ---        & Area ($n$+1)& 0.80$^\ast$&0.82$^\ast$\\
 \hline
15--22 & $\frac{ {\rm Area} \times \langle {\rm Tilt}\rangle } { \langle {\rm Latitude} \rangle}$  ($n$) & Polar field ($n$)& 0.94$^\ast$& 0.94$^\ast$\\
15--22  & ~~~ ---       & Area ($n$+1)&  0.87$^\ast$& 0.86$^\ast$\\
 \hline
\end{tabular}
\label{table1}
\end{center}
\tablecomments{
The value $r_{\rm all}$ represents the correlation for cycles mentioned in the first column, while $r_{\rm all}$ excludes Cycle 19--20 pair. The values with $\ast$ in the superscript are statistically significant having $p$ value $< 0.05$.
}
\end{table}

\begin{table*}
\begin{center}
\caption{ 
The same as \Tab{table1}, but here the analyses are extended till Cycle 24.
}
\begin{tabular}{cllclll}     % define the column alignment
                           % l: left, c: center, r: right
\hline                     % horizontal line
Cycles & \multicolumn{2}{c}{Quantities} & $r_{\rm all}$ & $r^{-}_{19/20}$ & $r^{-}_{23/24}$ & $r^{-}_{19/20, 23/24}$\\
\hline                     % horizontal line
15--24 & Area ($n$) & Polar field ($n$) & 0.38$^{~}$ & 0.68$^\ast$& 0.43$^{~}$& 0.80$^\ast$\\
12--24& ~~~ ---    & Area ($n+1$)      & 0.57$^\ast$& 0.67$^\ast$& 0.69$^\ast$ &0.85$^\ast$\\
\hline
15--24 & $\frac{ {\rm Area} }{ \langle {\rm Latitude} \rangle}$ ($n$)& Polar field ($n$)& 0.48$^{~}$& 0.74$^\ast$& 0.53$^{~}$&0.86$^\ast$\\
12--24 & ~~~ ---        & Area ($n$+1)& 0.62$^\ast$ &  0.71$^\ast$& 0.75$^\ast$ & 0.88$^\ast$\\
 \hline
15--24& ${\rm Area} \times \langle {\rm Tilt}\rangle$ ($n$)& Polar field ($n$)& 0.58$^\ast$ &  0.60$^\ast$&0.80$^\ast$&  0.87$^\ast$\\
15--24 & ~~~ ---        & Area ($n$+1)& 0.27$^{~}$&  0.26$^{~}$&0.80$^\ast$&  0.82$^\ast$\\
 \hline
15--24 & $\frac{ {\rm Area} \times \langle {\rm Tilt}\rangle } { \langle {\rm Latitude} \rangle}$  ($n$) & Polar field ($n$)& 0.68$^\ast$ &  0.66$^{~}$& 0.91$^\ast$ &  0.92$^\ast$\\
15--24 & ~~~ ---       & Area ($n$+1)&  0.32$^{~}$&  0.30$^{~}$& 0.87$^\ast$ &  0.86$^\ast$\\
 \hline
\end{tabular}
\label{table2}
\end{center}
\tablecomments{
For the polar field, we used the $A(t)$ index for Cycles 15--22 and for Cycles 23--24, the WSO polar field, scaled appropriately to make the value using linear regression.  $r_{23/24}^-$ represents the correlation value excluding the Cycle 23--24 pair and $r_{19/20, 23/24}^-$ represents the same but excluding both 19--20 and 23--24 pairs. 
}
\end{table*}

So we look for the third candidate, which is the tilt angle. 
\citet{Jiao21} provided the average tilt angle of past cycles, starting from solar cycle 15 from Mount Wilson Observatories (we took their values from Table 8 with $\Delta s \ge 2.5^\circ$).
Using their data, we multiply the total area by the cycle-averaged tilt angle and obtain the correlation. As seen in \Tab{table1}, the correlations for both the polar field and the amplitude of the next sunspot cycle are remarkably high, even including the 19--20 cycle pair. Removing this pair increases the correlation value only marginally; also see \Fig{fig:latvstilt}(b), how the data point for 19--20 cycle pair moved closer to the linear trend which was an outlier in previous correlations (e.g., \Fig{fig:latvstilt}(a) or \Fig{fig:quench}(a)) and the reduction of $p$ and $\chi^2$ values.  This suggests that the tilt angle, in combination with flux, plays a crucial role in determining the amplitude of the polar field, particularly for Cycle 19, which eventually produces a weak cycle (Cycle 20).
Finally, when we include both the mean latitude and the tilt angle, we expectedly observe the highest correlations as shown in \Fig{fig:latvstilt}(c) and the last two rows of \Tab{table1}. Even the elimination of 19--20 pair does not alter the correlation. The correlation in the last row of \Tab{table1} agrees with the results of \citet{Das10}, who found a strong correlation between the strength of cycle multiplied by normalised mean area-weighted tilt angle versus the strength of the next cycle; also see \citet{Muno13} for the same correlation but with the polar faculae (higher correlations in our case could be due to a slightly different way of computing the parameters and the use of updated data sets). However, they did not identify the importance of the tilt angle over the latitude.

The above discussion was based on only Cycles 15--22 for which we have a uniform proxy of the polar field ($A(t)$ index). Now we try to extend the above discussion to the maximum possible cycles. For recent cycles, we have measurements of the polar field. So for Cycle 23--24, we use the WSO polar field considering with an appropriate scalling factor (fixed based on the regression relation) to bring it into the $A(t)$ index scale. The correlations of various quantities with the polar field for Cycles 15--24 are given in \Tab{table2}.  
For area and latitude, we anyway have the record from cycle 15, while for tilt angle, we take its value for Cycles 23--24 from Table 6 of unbinned data with $\Delta s \ge 2.5^\circ$ of \citet{Jiao21}. 

From \Tab{table2} we observe a similar conclusion as obtained for Cycles 15--22 as presented in \Tab{table1} except for the correlation with the tilt angle, namely, $\rm {Area} \times \langle \rm{Tilt} \rangle $ with the polar field or with the area of the next cycle.  Even excluding Cycle 19--20 pair, does not increase the correlation; however, excluding Cycle 23--24 pair increases the correlation abruptly. This implies that it is the Cycle 23--24 pair, for which the $average$ tilt angle in combination with the BMR area cannot explain the polar field of Cycle 23 (or the strength of Sunspot Cycle 24). It was already explored by \cite{JCS15} that a number of large $wrongly$ tilted (anti-Hale or anti-Joy)  BMRs emerged at low latitudes (which are highly efficient in generating the polar field) caused the weak polar field in Cycle 23; also see \citet{PN25} for modelling Cycles 16-24 using tilt scatter. This suggests that the average tilt (over the whole cycle) may not fully capture the effect of tilt fluctuations, and the individual measurement of tilt of each BMR from magnetic field data is essential. 
   
\section{Conclusion} %%%%%%%%%%%%%%%%%%%%%%%%%%%%%%%%%%%%%%%%
\label{sec:conclusion} 
Nonlinearity and stochastic fluctuations in the poloidal field generation through the decay of sunspots are the two main components for causing cycle-to-cycle variations and modulations in the solar cycle \citep{Karak23}. 
Latitude and tilt quenchings are the two potential candidates for the nonlinearity in the polar field generation, as offered in recent theoretical studies guided by observations \citep{Das10, jha20, Petrovay20, J20, Kar20, yeates25}. And stochastic fluctuations in the BMR properties are identified as the cause of the polar field variation \citep[e.g.,][]{Ca13, JCS14, Nagy17, KM17, pal23, BKK23, Kumar24}. 
However, the relative contributions of these nonlinearities were not disentangled and how the BMR properties predict the polar field and thus the amplitude of the next cycle were also not identified in the previous studies.

By analysing the observed sunspot tilts and latitudes and the polar field proxy data in the present paper, we show that, for the majority of cycles, the polar field generated in a cycle (and thus the strength of the next cycle) is linearly correlated with the total area in sunspots. The stochastic fluctuations in sunspot properties introduce a scatter in this linear dependency. The systematic variation in sunspot properties, such as strong cycles producing spots at high latitudes and spots with weak tilt angles, also alters the polar field. The mean latitude and tilt angle, combining with the total sunspot area of a cycle, form the precursor of the polar field, and thus the next cycle strength. 
We show that for strong cycles, the nonlinearity plays an essential role to cause a decrease in the polar field, which consequently reduces the next sunspot cycle strength (e.g., Cycle 19--20 and 4--5 pairs). 
For solar cycle 19, which followed a considerably weak cycle, the tilt quenching plays a dominant role over latitude quenching. 
Due to the unavailability of sunspot latitude and tilt for the cycle pair 4--5, it remains to be explored what caused solar cycle 5: Is it the tilt or latitude quenching?

For Cycles 15--22, for which we have the data of uniform polar field proxy, we show that the tilt angle variation dominates the latitude variation, 
which contrasts with \citet{yeates25}.
Combining the WSO polar field for Cycles 23--24 with $A(t)$ index, we find similar conclusion to hold for Cycles 15--24, except for the Cycle 23--24 pair. For Cycle 23, the average tilt angle may not determine the polar field, a few wrongly tilted BMRs appearing at low latitudes as suggested by \citet{JCS15, Nagy17, KM18} can alter the polar field significantly without changing the cycle-average tilt much. Other mechanisms, such as changes in inflows \citep{Jiang10} or meridional flow \citep{Kar10, nandyNAT}, may also have some effect in determining the polar field of Cycle 23.  

A significant correlation between the sunspot cycle strength (or the polar field) of $n$ and the cycle strength of $n+1$ observed during Cycles 15--23, except for the pair 19/20 shows that the solar cycle indeed has a weak memory that extends multiple cycles \citep{Kumar21b}, as a robust outcome of the solar dynamo operating at near-critical regime \citep{Wavhal25, V23}. However, this memory is insufficient to predict the amplitudes of multiple following cycles beyond one because the variations in the BMR properties, primarily in the tilt angle, can alter the polar field.   
%%%%%%%%%%%%%%%%%%%%%%%%%%%%%%%%%%%%%%%%%%%%%%%%%%%%%%%%%%%%%%%%%%%%%%%%%%%

\begin{acknowledgments}
\textbf{Acknowledgements} 
The authors appreciate the comments and corrections made by the anonymous referee which helps us to improve the quality of the paper. B.B.K. acknowledges the financial support from the Indian Space Research Organisation (project no. ISRO/RES/RAC-S/IITBHU/2024-25).
\end{acknowledgments}

\begin{acknowledgments}
\textbf{Author Contribution} All the analyses of the data and computations were conducted by B.D. under the guidance of B.B.K. and A.S. A.S. crosschecked the results and prepared the figures. B.B.K. wrote the manuscript, while A.S. and B.D. reviewed and edited it. 
\end{acknowledgments}

\begin{acknowledgments}
\textbf{Data Availability} Sunspot number data were obtained from WDC-SILSO, Royal Observatory of Belgium, Brussels, DOI: \url{https://doi.org/10.24414/qnza-ac80} \citep{SILSO_Sunspot_Number}.
Sunspot area data were obtained from the calibrated sunspot dataset of \citet{Mandal2020}. The sunspot tilt information was taken from the Mount Wilson Observatory data provided in \citet{Jiao21}.
\end{acknowledgments}

\begin{acknowledgments}
\textbf{Conflict of interest} The authors declare that they have no conflicts of interest.
\end{acknowledgments}

\bibliographystyle{yahapj}
\bibliography{bidisha}

\end{document}